\documentclass[shortnote]{jpsj3_arXiv}
\usepackage{txfonts}
\usepackage[dvipdfmx]{graphicx}
\usepackage{mediabb}
\usepackage{color}

\title{Improved frequency measurement of the $^1S_{0}$-$^3P_{0}$ clock transition in $^{87}$Sr using the Cs fountain clock at NMIJ as a transfer oscillator}

\author{Takehiko Tanabe$^1$\thanks{t.tanabe@aist.go.jp}\footnote[3]{These two authors contributed equally to this work.}, Daisuke Akamatsu$^{1\ \ddag}$, Takumi Kobayashi$^1$, Akifumi Takamizawa$^1$, Shinya Yanagimachi$^1$, Takeshi Ikegami$^1$, 
Tomonari Suzuyama$^1$, Hajime Inaba$^1$, Sho Okubo$^1$, Masami Yasuda$^1$, Feng-Lei Hong$^{1,2}$, Atsushi Onae$^1$, and Kazumoto Hosaka$^1$\thanks{kazu.hosaka@aist.go.jp}}
\inst{$^1$National Metrology Institute of Japan (NMIJ), National Institute of Advanced Industrial Science and Technology (AIST), Umezono 1-1-1, Tsukuba 305-8563, Japan \\
$^2$Department of Physics, Graduate School of Engineering, Yokohama National University, 79-5 Tokiwadai, Hodogaya-ku, Yokohama 240-8501, Japan} 

\abst{We performed an absolute frequency measurement of the $^1S_{0}$-$^3P_{0}$ transition in $^{87}$Sr with a fractional uncertainty of $1.2 \times 10^{-15}$, which is less than one third that of our previous measurement.
A caesium fountain atomic clock was used as a transfer oscillator to reduce the uncertainty of the link between a strontium optical lattice clock and the SI second.
The absolute value of the transition frequency is 429 228 004 229 873.56(49) Hz.}

\begin{document}
\maketitle
Recently, some optical clocks have reached the $10^{-18}$ level\cite{Ushijima2015,Nicholson2015} in both uncertainty and stability, and these values surpass the caesium fountain microwave primary standards used to realise the SI unit of time.
The high-performance of the optical clocks means that the scientific community is discussing a re-definition of the second.
Therefore, there is a need for the metrology community to make a strenuous effort to determine the absolute frequencies of the optical frequency standards in relation to the current primary frequency standards,
so that the length of one second remains unchanged after the re-definition.

At the National Metrology Institute of Japan (NMIJ) we have developed atomic clocks based on optical transitions in an ensemble of neutral atoms trapped in Stark-shift-free optical lattices\cite{Kohno2009,Yasuda2012,Akamatsu2014}, which are called optical lattice clocks\cite{Takamoto2005}.
In 2014, we measured the frequency of the $^1S_{0}$-$^3P_{0}$ clock transition in $^{87}$Sr\cite{Akamatsu2014}.
At that time the uncertainty of the absolute frequency ($3.7 \times 10^{-15}$) was mainly limited by the uncertainty of a comparison with NMIJ coordinated universal time (UTC(NMIJ)),
because we do not know how UTC(NMIJ) changed during the exact measurement time, even though the frequency differences between UTC(NMIJ) and the International Atomic Time (TAI) averaged over five-day campaign periods are provided in Circular T,
which is a monthly report produced by the ``Bureau International des Poids et Mesures (BIPM)''.
There is an effective and practical averaging time for each different type of clock that allows it to perform with good stability and a small uncertainty.
Optical lattice clocks and microwave clocks differ in this respect.
This is one of the difficulties involved in achieving accurate frequency comparisons of clocks.
Although optical lattice clocks have excellent short-term frequency stabilities, operating them continuously for five days would be a very difficult task owing to the problem of lasers losing lock, which introduces distributed dead time into frequency comparisons.
To minimize the uncertainty of the frequency comparison with UTC(NMIJ), it is important to understand the effect of the distributed dead time and the behaviour of UTC(NMIJ).
\begin{figure}
\begin{center}
  \includegraphics[scale=0.3,clip]{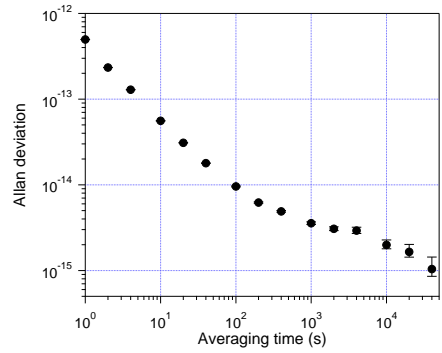}
  \caption{A fractional Allan deviation calculated from the phase difference between hydrogen masers. One of the masers has been used as the reference signal of UTC(NMIJ).}
\end{center}
\end{figure}
In this paper, we carefully evaluate the uncertainties of the link between our Sr optical lattice clock and TAI via UTC(NMIJ) using a caesium fountain atomic clock located at NMIJ (NMIJ-F2)\cite{Takamizawa2014} as a transfer oscillator.
In this way, we reduced the final uncertainty to one third that of our previous measurement\cite{Akamatsu2014}.

The experimental set-up of our Sr optical lattice clock has been described in detail elsewhere\cite{Akamatsu2014}, so in this paper we mention only the main modifications.
For an absolute frequency measurement, a fibre-based optical frequency comb referenced to UTC(NMIJ) is used.
A hydrogen maser has been used as the reference signal of UTC(NMIJ). Figure 1 shows a fractional Allan deviation calculated from the phase difference between the hydrogen maser and another maser, which was observed during the measurement period described later.
When we measure the clock transition frequency in Sr, we cannot compare it directly with the NMIJ-F2, because NMIJ-F2 uncertainty evaluations have not been completed.
Using the NMIJ-F2 as a transfer oscillator that operated continuously during the measurement period, we could consistently monitor the time development of UTC(NMIJ), and so could compensate for the drift of the reference for the comb.
NMIJ-F2 can be compared with UTC(NMIJ) through a cryogenic Sapphire oscillator (CSO), which is applied to operate NMIJ-F2 via frequency synthesis process for generating the 1 GHz reference frequency.
Phase comparison data regarding the transfer oscillator was collected with another data acquisition system that was synchronised to within a second with that for the Sr optical lattice clock, and the comb system and the data were subsequently combined.
The experimental parameters for the Sr optical lattice clock in this measurement, such as magnetic field and laser power, are the same as those that we used for our previous measurement.
Using these experimental conditions, we recently evaluated the frequency correction of the clock transition as $37.0 \times 10^{-16}$ with a relative uncertainty of $3.8 \times 10^{-16}$\cite{Akamatsu2014}. 

Clock transition frequency data was collected for more than 70000 s over a five-day measurement campaign during the period MJD 57184 to MJD 57199, which corresponds to three of the five-day intervals reported in Circular T.
There is a need for using a proper length of the averaging time to compare the clocks, because the interrogation cycle of the Sr optical lattice clock is different from that of NMIJ-F2. In this experiment, 40 s averaging time, which is close to a common multiple of the both cycles, is selected.
Figure 2(a) shows the entire counting record with a 40 s gate time, after removing the data corresponding to the times when any lasers lost lock,
and also the times when tracking oscillators used for beat frequency measurements lost lock.
\begin{figure}
\begin{center}
  \includegraphics[width=\columnwidth,clip]{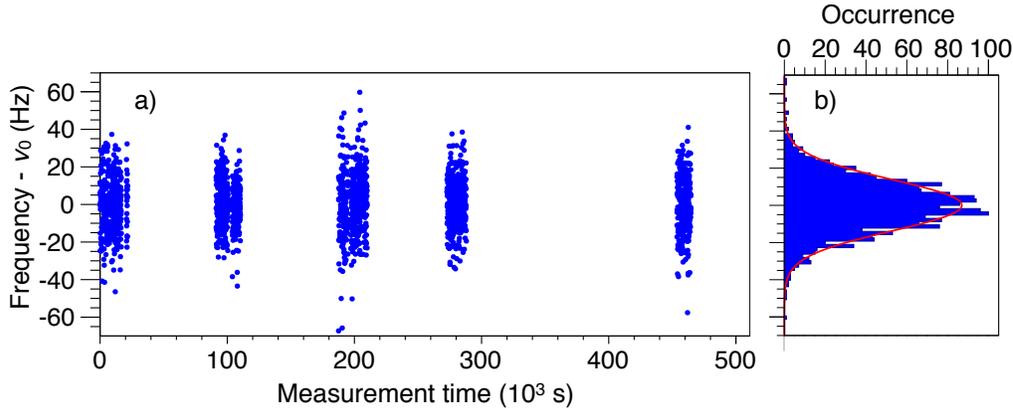}
  \caption{Measured frequencies of the Sr clock transition compared with NMIJ-F2. a) Measurement data point as a function of time; each data point corresponds to a 40 s average. The offset frequency $\nu_{0}$ is the BIPM recommended frequency value.\cite{BIPM}. b) A histogram of the frequency measurement shown in a).}
\end{center}
\end{figure}
In this figure, the above-mentioned systematic shifts for the Sr optical lattice clock are corrected and the drift of UTC(NMIJ) is compensated for by using the frequency comparison data between UTC(NMIJ)
and the NMIJ-F2, assuming white frequency noise for NMIJ-F2. 
Frequency noise of NMIJ-F2 had been evaluated by comparing with CSO and we have observed the frequency stability of $1.3 \times 10^{-13}/\tau^{1/2}$, where $\tau$ is the averaging time.

A histogram of the counting record is shown in Fig. 2(b), plotted as an offset from the BIPM recommended value: $\nu_{0}=429\ 228\ 004\ 229\ 873.4$ Hz\cite{BIPM}.
The solid red line shows a Gaussian fit to the data, indicating that the mean frequency is 0.12(34) Hz with a full width at half maximum of 33 Hz.
The standard error of the mean of the counting records, which should include all link errors between the Sr optical lattice clock and NMIJ-F2, is counted in the final measurement uncertainty budget as a statistical uncertainty.
Environmental condition in the laboratory is fairly good, so that constant frequency shifts induced by environmental perturbations such as temperature drift on the optical path including uncompensated fibres should be significantly small.

When comparing the Sr clock frequency with that of NMIJ-F2, we need to think about the dead time uncertainty, because the Sr optical lattice clock was operated for only several hours a day. 
When we assume white frequency noise for NMIJ-F2, we estimate the dead time uncertainty to be $4.7 \times 10^{-16}$
for the total measurement time of 71000 s over a five-day measurement campaign during the period MJD 57184 to MJD 57199, which corresponds to three of the five-day intervals reported in Circular T.\cite{Parker1998}
The correction between NMIJ-F2 and UTC(NMIJ) is calculated to be $41.7(2.0) \times 10^{-16}$ using the comparison data for the fifteen days, where the uncertainty is based on the standard deviation of the mean.
The correction of the link between UTC(NMIJ) and TAI can be calculated from the Circular T 330\cite{CircularT330} and was found to be $-32.4(3.7) \times 10^{-16}$ for the campaign periods\cite{Parker2007}.
A correction of $-8.3(1.8) \times 10^{-16}$ between TAI and SI was also obtained from Circular T 330\cite{CircularT330}.
There is a need to estimate the dead time uncertainty for TAI and SI comparison, because the correction value and the uncertainty shown in Circular T 330 are calculated based on data for 30 days.
It is estimated to be $3.6 \times 10^{-16}$ for the fifteen days\cite{Parker1998}.
These corrections and their standard uncertainties, which should be considered for the clock transition frequency presented in Fig 2, are summarized in Table 1.

In conclusion, we have measured the absolute frequency of the $^1S_{0}$-$^3P_{0}$ clock transition in $^{87}$Sr with a final fractional uncertainty of $1.2 \times 10^{-15}$,
which is mainly limited by the standard error of the mean of the counting records when the Sr optical lattice clock is compared with NMIJ-F2.
The measured frequency is $429\ 228\ 004\ 229\ 873.56(49)$ Hz.
The result of this work is in good agreement with the recommended value and previously measured values by several different groups including our previous result.\cite{Akamatsu2014,Campbell2008,Baillard2008,LeTargat2013,Hong2009,Falke2011,Falke2014,Hachisu2014}.
The measured value will be reported to the ``Comit\'e International des Poids et Mesures (CIPM)'' and will contribute to discussions regarding the updating of the recommended frequency value of the clock transition in $^{87}$Sr.
\begin{table}
\caption{Frequency corrections applied to the data of the Sr clock transition frequency compared with NMIJ-F2 (shown in Fig. 2) and their standard uncertainties. Since the systematic shifts for the Sr optical lattice clock are already corrected in Fig. 2, it is written in parenthesis.}
\label{t1}
\begin{center}
\begin{tabular}{lcc}
\hline
\multicolumn{1}{c}{Effect} & \multicolumn{1}{c}{Correction $(10^{-16})$} & \multicolumn{1}{c}{Uncertainty $(10^{-16})$} \\
\hline
Sr systematics            &                  $(37.0)$ & $3.8$   \\
Sr$-$NMIJ-F2 statistics   &                           & $7.9$   \\
Dead time in Sr$-$NMIJ-F2 &                           & $4.7$   \\
NMIJ-F2$-$UTC(NMIJ)       &                    $41.7$ & $2.0$   \\
UTC(NMIJ)$-$TAI           &                   $-32.4$ & $3.7$   \\
TAI$-$SI                  &                    $-8.3$ & $1.8$   \\
Dead time in TAI$-$SI     &                           & $3.6$   \\
\hline
Total                     &                     $1.0$ & $11.5$  \\
\hline
\end{tabular}
\end{center}
\end{table}

\end{document}